\documentclass[conference]{utils/IEEEtran}

\usepackage[utf8]{inputenc}
\usepackage{cite} 
\usepackage[pdftex]{graphicx} 
\usepackage{amsmath} 
\usepackage{array} 
\usepackage{url} 
\usepackage{float}
\usepackage{subfig}

\graphicspath{{inc/graphics/}}

\IEEEoverridecommandlockouts
\IEEEpubid{\makebox[\columnwidth]{978-1-5386-7848-0/18/\$31.00~\copyright~2018 IEEE \hfill} \hspace{\columnsep}\makebox[\columnwidth]{ }}

\begin{document}

\title{
  Exploratory Data Analysis of a Network Telescope Traffic
  and Prediction of Port Probing Rates
}

\author{
  \IEEEauthorblockN{
    Mehdi Zakroum\IEEEauthorrefmark{1},
    Abdellah Houmz\IEEEauthorrefmark{1}\IEEEauthorrefmark{3},
    Mounir Ghogho\IEEEauthorrefmark{1},
    Ghita Mezzour\IEEEauthorrefmark{1},
  }
  \IEEEauthorblockN{
    Abdelkader Lahmadi\IEEEauthorrefmark{4},
    Jérôme François\IEEEauthorrefmark{4},
    and Mohammed El Koutbi\IEEEauthorrefmark{3}
  }
  \IEEEauthorblockA{\IEEEauthorrefmark{1}International University of Rabat, TICLab, Morocco}
  \IEEEauthorblockA{\IEEEauthorrefmark{3}Mohammed V University of Rabat, ENSIAS, Morocco}
  \IEEEauthorblockA{\IEEEauthorrefmark{4}Université de Lorraine, CNRS, Inria, LORIA, F-54000 Nancy, France}
  \IEEEauthorblockA{
    \texttt{\{mehdi.zakroum,abdellah.houmz,mounir.ghogho,ghita.mezzour\}@uir.ac.ma}\\
    \texttt{\{abdelkader.lahmadi,jerome.francois\}@inria.fr}\\
    \texttt{elkoutbi@ensias.ma}
  }
}

\maketitle

\IEEEpubidadjcol

\begin{abstract}
Understanding the properties exhibited by large scale network probing traffic
would improve cyber threat intelligence. In addition, the prediction of probing
rates is a key feature for security practitioners in their endeavors for making
better operational decisions and for enhancing their defense strategy skills.
In this work, we study different aspects of the traffic captured by a /20
network telescope.
First, we perform an exploratory data analysis of the collected probing
activities. The investigation includes probing rates at the port level, 
services interesting top network probers and the distribution of probing 
rates by geolocation.
Second, we extract the network probers exploration patterns. We model these
behaviors using transition graphs decorated with probabilities of switching
from a port to another.
Finally, we assess the capacity of Non-stationary Autoregressive and Vector
Autoregressive models in predicting port probing rates as a first step towards
using more robust models for better forecasting performance.

\end{abstract}

\vspace{5mm}


\section{Introduction}
New cyber threat vectors and vulnerabilities are constantly emerging with the
evolution of technology.  Attackers commonly scan networks to find vulnerable
devices which can be used for malicious intents.  One of the major attacks
happened in 2016 is the Dyn DDoS attack.  The attackers used botnets of
vulnerable devices as a primary source of their DDoS traffic generation, making
leading internet platforms unavailable for a large number of users.

Improving our knowledge on scan activities will help to prevent cyber attacks
through early detection, and in general, to enhance security policies.  Many
causes can trigger scan campaigns such as vulnerability disclosure, worm spread
and zero days.  Generally, such malicious traffic is hidden by a large amount
of legitimate traffic, making it complex to be identified by internet service
providers and network security operators to protect target users.

A passive approach for identifying network probing activities are network
telescopes, also known as darknets.  A network telescope is a sensor logging
the traffic received by a set of passive unallocated network addresses.
Therefore, the traffic received by the network telescope is considered
suspicious, requiring thus to be examined.

To collect such traffic, we use a network telescope hosted at INRIA
Nancy-Grand Est consisting of nearly 4096 IPv4 addresses. By analyzing the
collected data, we aim to answer the following questions:
\begin{itemize}
  \item What are the most targeted services? What are the services targeted by
    the top network probers?
  \item How network probers are exploring the target network? How to model
    these probing activities?
  \item Can we predict probing rates of the targeted services?
\end{itemize}

The remainder of this paper is structured as follows. The next section provides
a review of the related work to this study. Section~\ref{sec:eda} presents an
exploratory analysis of the darknet traffic. In
section~\ref{sec:probing_patterns}, we identify the attackers probing patterns.
Finally, in section \ref{sec:probing_rates}, we explore the capacity of
Non-stationary Autoregressive and Vector Autoregressive models to forecast the
probing rates at the port level.

\label{sec:introduction}

\section{Related Work}
%
%
Reconnaissance is the first phase in the cyber kill chain, where the attacker
scans the target infrastructure looking for vulnerabilities. A more generic
approach for finding vulnerable devices consists of scanning the whole IPv4
address space, including network telescopes. Many studies leverage
the traffic captured by the latter to study different aspects of
probing activities.
Durumeric et al.  \cite{durumeric_internet-wide_2014} studied the traffic
acquired by a large network telescope consisting of 5.5M IP addresses. The
study includes the origin of scans, the targeted services by network
probers and the effect of vulnerability publication on probing activities.
Bou-harb et al. \cite{bou-harb_statistical_2013} used a probabilistic
and statistical approach to identify the origin of the probing activities:
whether they are generated by scanning tools or by worms and botnets. They also
studied whether probing activities are random or they exhibit specific patterns.
Eto et al. \cite{hutchison_proposal_2009} proposed a method to extract the
features of scanning malwares based on the oscillation of destination IP
addresses in the captured scan packets.
Li et al. \cite{li_honeynet-based_2008} proposed a general framework that
identifies scanning events and analyzing methods used by botnets in probing
compaigns. They applied their framework to extract the scanning characteristics
of a set of 6 botnets.
Papale et al. \cite{dainotti_analysis_2012} analyzed a 12-day world-wide cyber
scanning campaign targeting VoIP (SIP) servers caught by a /8 darknet. They
found that the origin is the Sality botnet which generated about 20 million
packet from roughly 3 million IP addresses.

Few studies explored the dependencies between targeted ports.
McNutt and Markus \cite{mcnutt_correlations_2005}
presented a method for detecting the start of anomalous port-specific activity
by recognizing deviation from correlated activities.
They found a high correlation between time series of flow counts on unassigned
or obsolete ports that do not have active services. Therefore, they can detect 
ports receiving anomalous activities.
In contrast to our work, they used in their study a traffic of an organization
network (not a darknet traffic), where the amount of benign traffic is large,
hiding thus malicious traffic.
%
Lagraa and François \cite{lagraa_knowledge_2017} inferred the dependency between
services using graph analysis. They proposed a graph-based model to discover
port scanning behavior patterns. They applied methods utilized for community
structure discovery in large graphs in order to identify clusters of ports.
Our work generalizes their approach: instead of constructing graphs for each
pair of source and destination IP addresses, our graphs aggregate the probing
activity by source IP address. This approach shows the general exploration
pattern followed by a network prober, regardless of the target host.

\label{sec:related_work}

\section{Exploratory Data Analysis}
\label{sec:eda}

  \subsection{Data Set}
  \label{subsec:data}
  The data we use is collected by a /20 network telescope. The traffic was
recorded from November 2014 until October 2017 and has a size of 2 TB. The
collected traffic consists of timestamped packet headers. We record for each
packet the source and the destination IP addresses, the source and the
destination ports and the packet's flags. Our study focuses on stateful
connections established from the source. Hence, we consider only packets with a
TCP SYN flag which count for approximately 4.5 billion packets.

  \subsection{Traffic by Port}
  \label{subsec:traffic_by_port}
\begin{figure}[!t]
  \centering
  \includegraphics[scale=0.31]{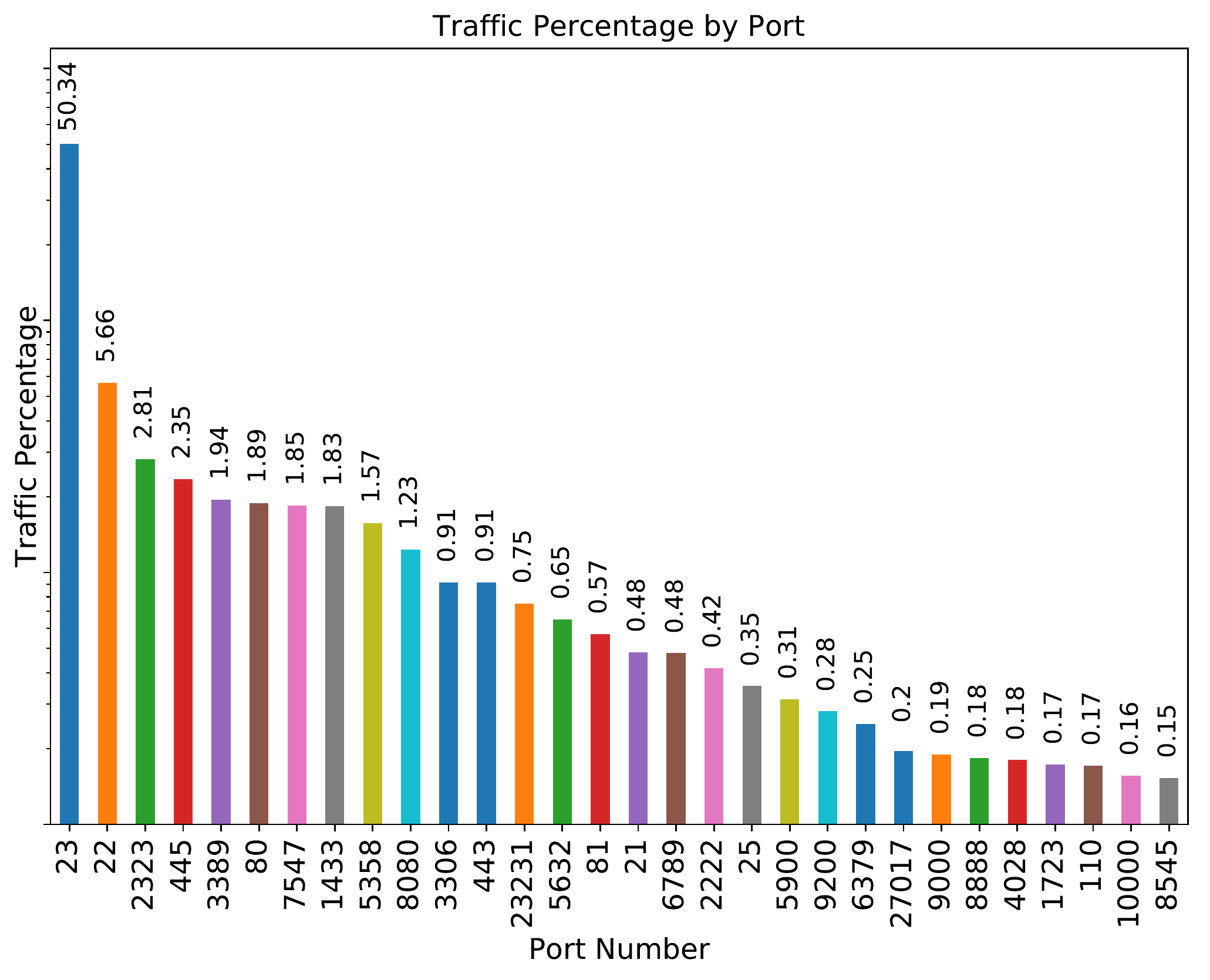}
  \caption{The 30 most targeted ports and their corresponding traffic
  percentage (the length of the bars are in log scale and the labels above the
  bars are the actual values)}
  \label{fig:ptp}
\end{figure}

\begin{figure}[!t]
  \centering
  \includegraphics[scale=0.31]{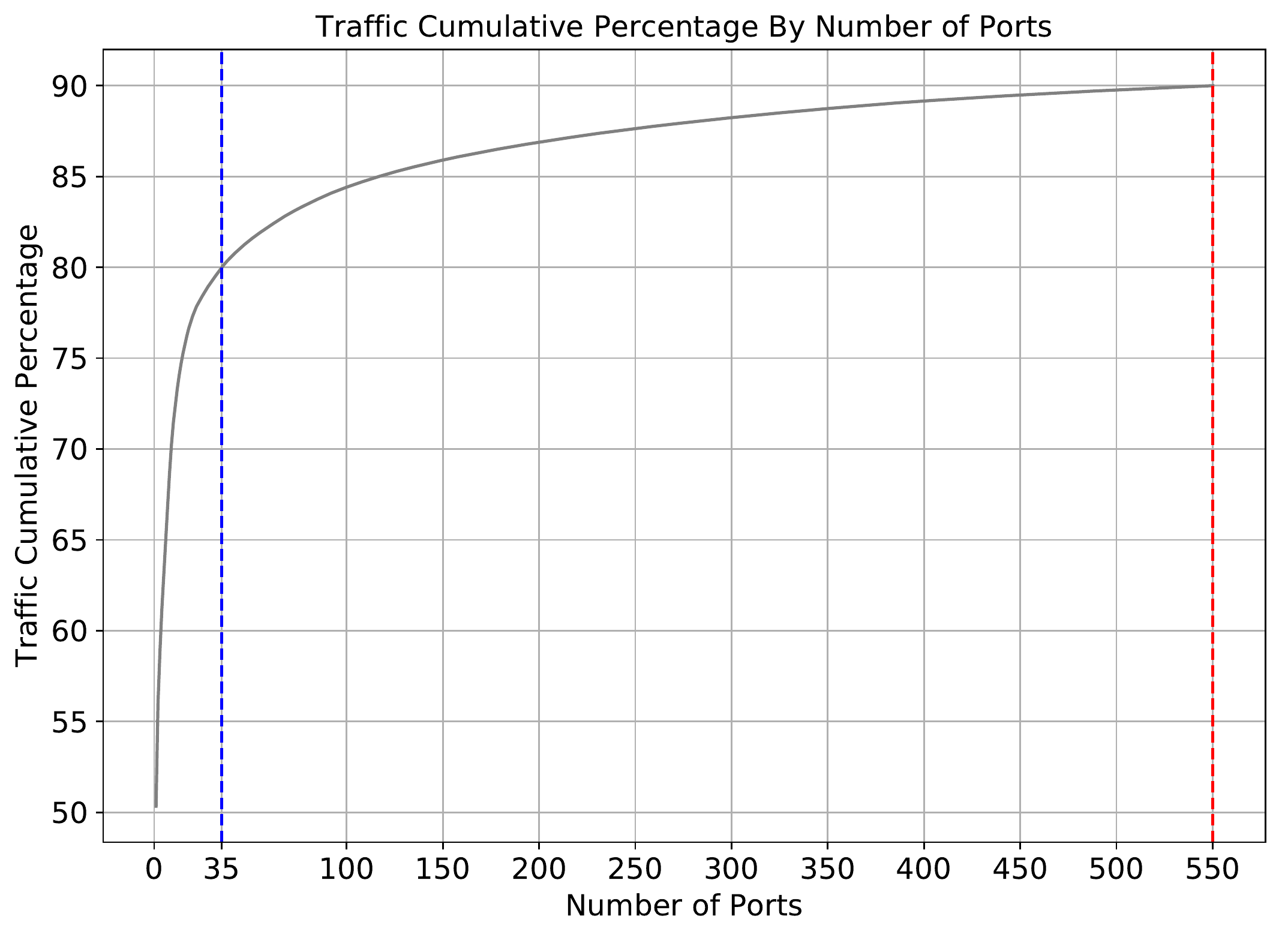}
  \caption{Traffic cumulative percentage by number of ports}
  \label{fig:nptcp}
\end{figure}

  We begin by extracting the traffic received by each port. That is, we aggregate
the received traffic by destination port and we count the number of TCP SYN
packets for each destination port. Figure~\ref{fig:ptp} shows the 30 most
targeted ports. We observe that the most targeted services are remote access
services, web servers, database management systems and some Microsoft services.
The network probers tend to use alternative ports in addition to the official
ones. The port 23 (telnet) generates more than 50\% of the traffic.
Figure~\ref{fig:nptcp} shows that among 65535 ports, only 35 generate 80\% of
the traffic and 550 generate 90\% of the traffic.

  \subsection{Top Network Probers' Interests}
  \label{subsec:top_network_probers_interests}
  The intent of a network prober might be manifested in the services he targets.
Knowing the ports that interest the top network probers determines services 
requiring particular security efforts. We consider as a top network prober one
maintaining an average probing rate higher than 150 TCP SYN packets per day. 
It is noteworthy that our definition of top network probers does not include 
probers performing distributed probing activities.

\begin{figure}[!t]
  \centering
  \includegraphics[scale=0.31]{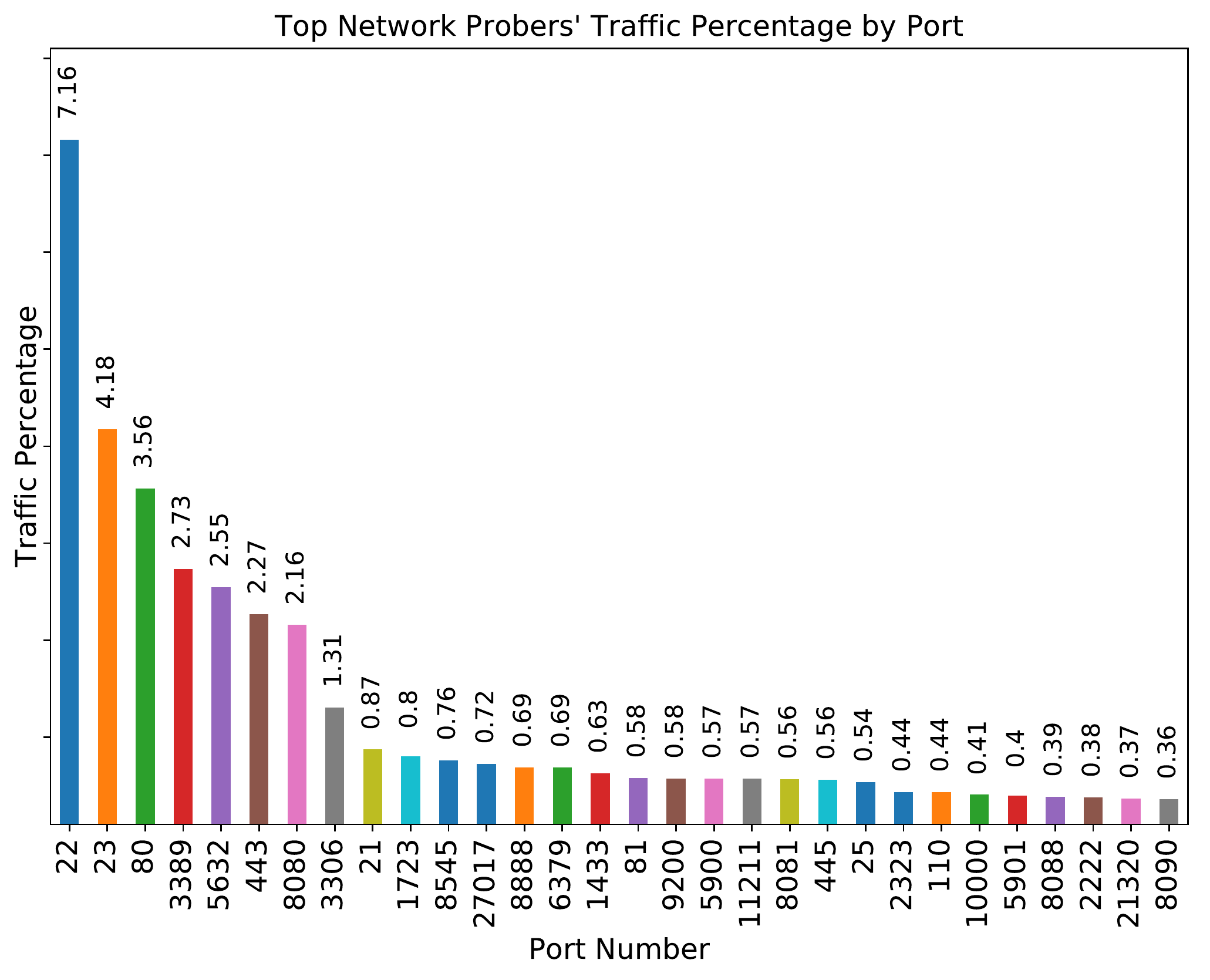}
  \caption{The 30 most targeted ports by top network probers (the numbers above
  the bars are the percentages of the traffic received by ports with respect to
  the total traffic generated by the top network probers)}
  \label{fig:tnptp}
\end{figure}

First, we count TCP SYN packets sent by more than 64 million source IP
addresses included in our data set. Then, for each top network prober (they are
nearly 1500), we extract the probing rates received by each port and we
aggregate the counts by port. Figure~\ref{fig:tnptp} shows the 30 most targeted
ports by top network probers. In contrast to the results in
Figure~\ref{fig:ptp}, top network probers focuse their probing activities on
the port 22 (SSH) rather than the port 23 (telnet).

  \subsection{Traffic by Country}
  \label{subsec:traffic_by_origin}
\begin{figure}[!t]
  \centering
  \includegraphics[scale=0.31]{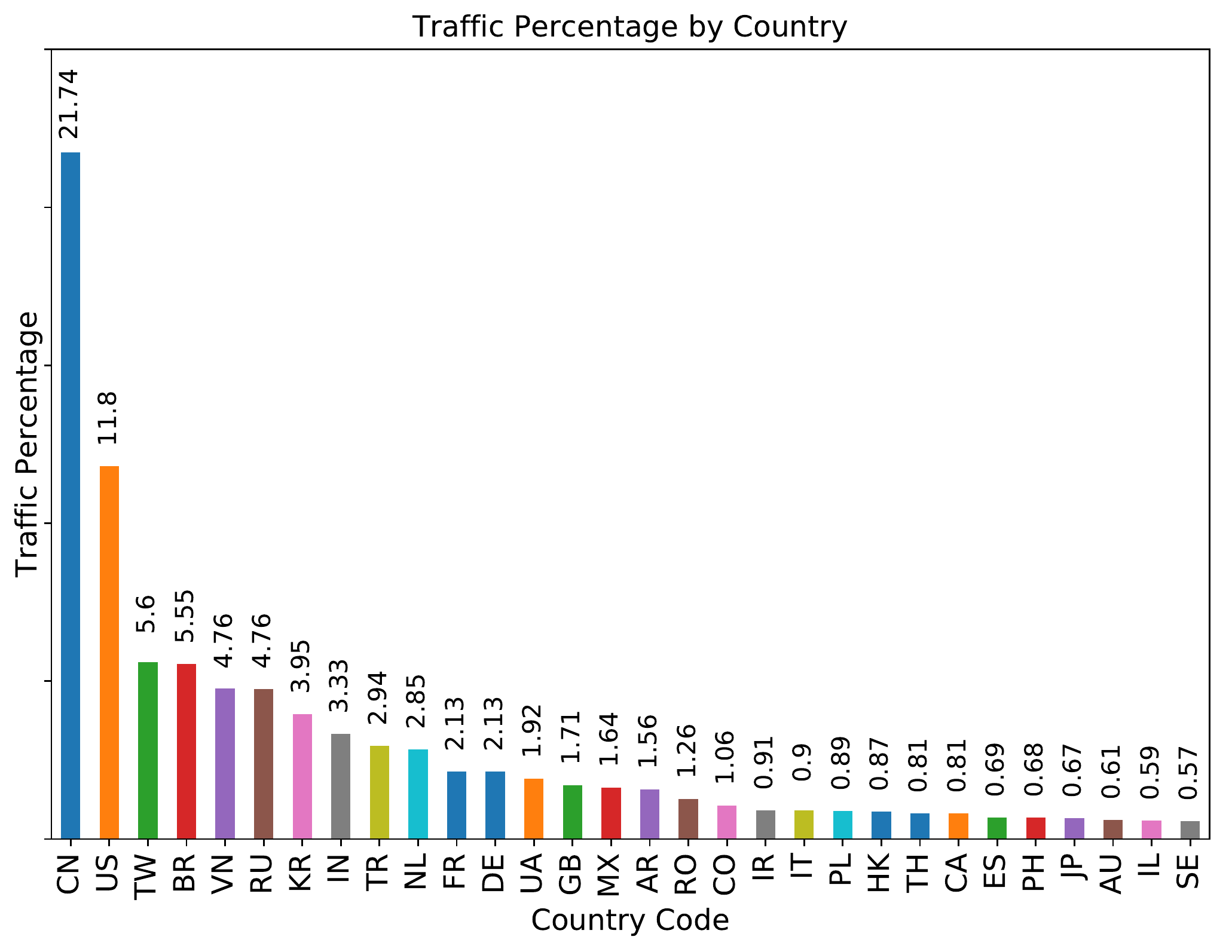}
  \caption{Top 30 countries and their corresponding traffic percentages}
  \label{fig:lptc}
\end{figure}

  The distribution of the received traffic by geolocation helps determining
how likely an occurred probing campaign is originating from a specific country.
We extract the total traffic caught by the network telescope by country. We
infer the country code from the source IP address using the DB-IP
database\footnote{URL: \url{https://db-ip.com/}}.  Figure~\ref{fig:lptc}
summarizes the received traffic by country code.

\section{Probing Patterns}
\label{sec:probing_patterns}
In this section, we model the behavior of network probers using
transition graphs. We assess the relationship between targeted services by
determining the probability of transition from a port to another, then, we
identify different network exploration patterns.

\subsection{Graph Modeling}

A network communication can be identified by a 5-tuple: source and
destination IP addresses, source and destination ports, and the transport
protocol. Our aim is to analyze for each network prober the exploration
behavior of the whole darknet. Hence, we take into account only two features:
the source IP address and the destination port.

To extract the graphs, we begin by aggregating the traffic generated by each
network prober. Then, we count the number of transitions from a destination
port to another by sequentially browsing the extracted traffic. These counts
are finally normalized in order to get the transition probabilities. It is to
note that the time dimension is omitted during this process.  

Formally, we extract for each source IP address \(i\) a transition graph
\(G_i(V_i, E_i)\), in which \(V_i\) is the set of destination ports targeted by
the network prober \(i\) and \(E_i\) represents the transition probabilities
between destination ports that are elements of \(E_i\). The association between
two elements \(p_a\) and \(p_b\) of \(E_i\) represents the probability that the
network prober \(i\) switches from \(p_a\) to \(p_b\).

\subsection{Extracted Graphs}

\begin{figure}[!t]
  \centering
  \subfloat[Sequential probing pattern (web services)]{
    \includegraphics[scale=.25]{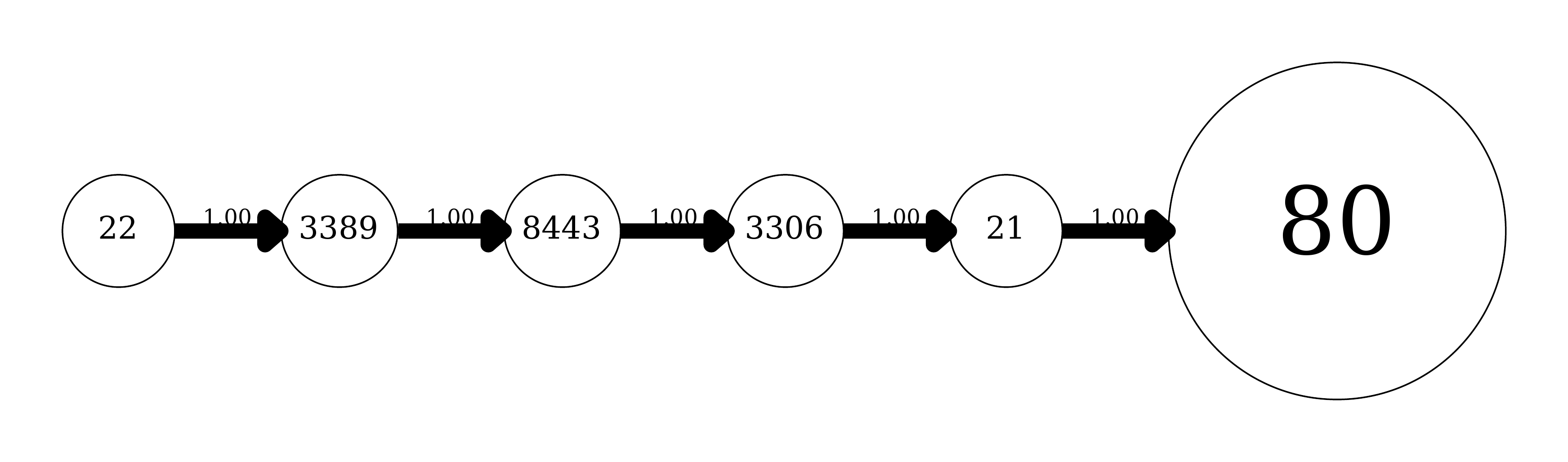}
    \label{fig:grapha}
  }
  \hfill
  \subfloat[Targeted probing pattern (MySQL service)]{
    \includegraphics[scale=.25]{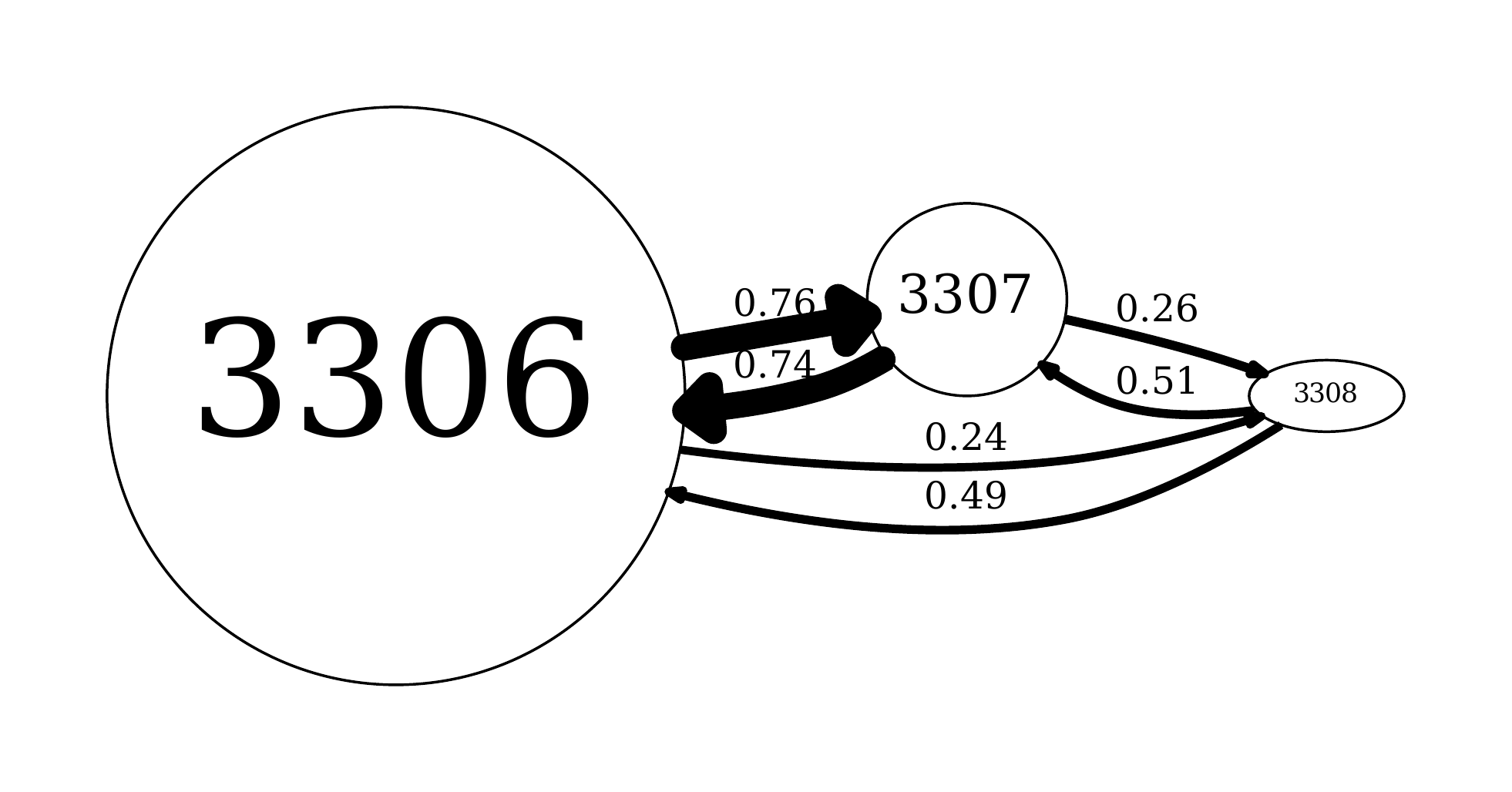}
    \label{fig:graphb}
  }
  \hfill
  \subfloat[Targeted probing pattern (remote access services)]{
    \includegraphics[scale=.25]{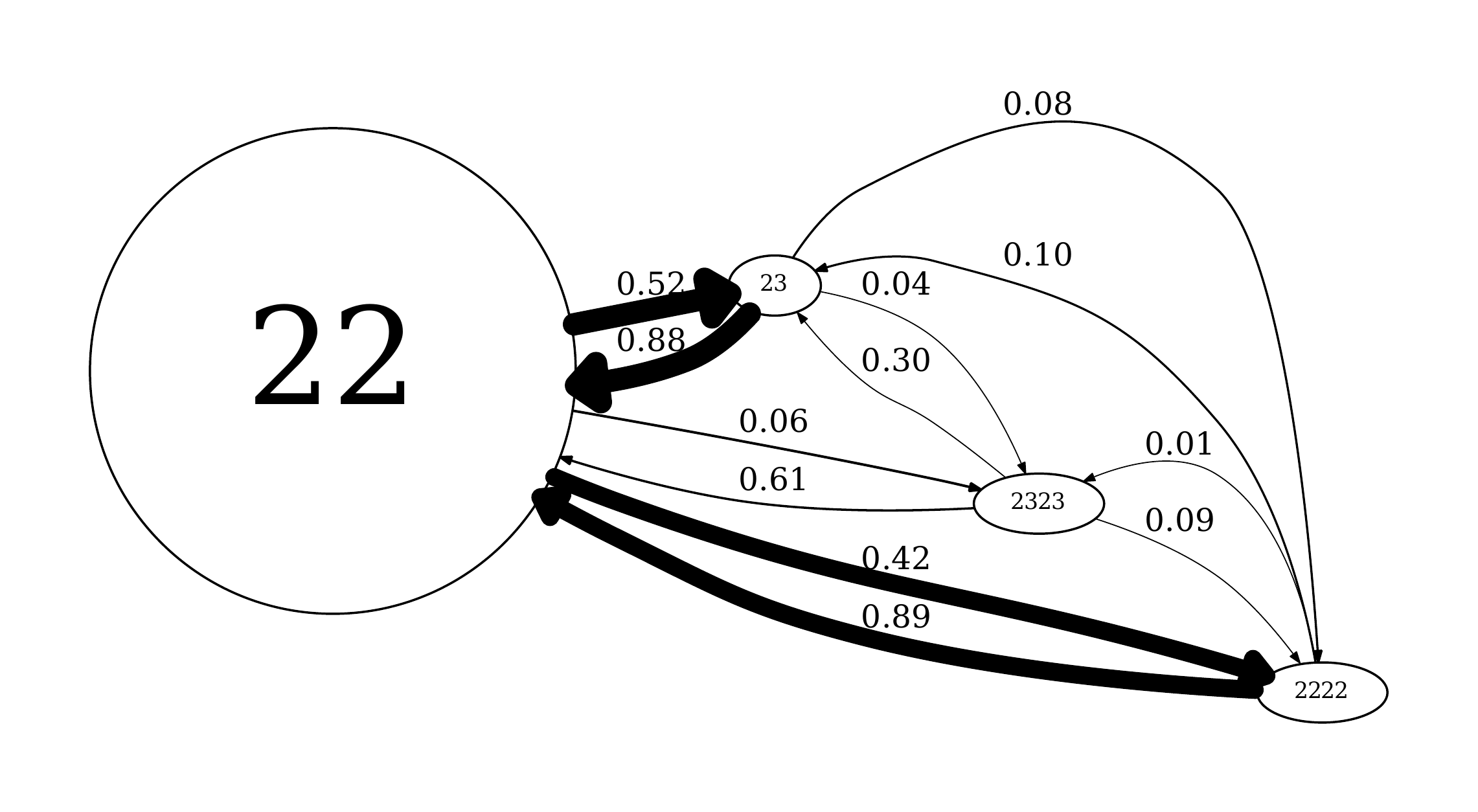}
    \label{fig:graphc}
  }
  \caption{A sample of transition graphs of 3 network probers. 
           The size of the vertices corresponds to the number of TCP SYN 
           packets received by the targeted port.}
  \label{fig:graph}
\end{figure}

Figure~\ref{fig:graph} shows a sample of transition graphs corresponding to 3
network probers.  Figure~\ref{fig:grapha} represents a network prober
sequentially targeting services typically deployed in web servers: SSH (22),
RDP (3389), MySQL (3306) and FTP (21), while focusing on the HTTP (80) server.
Figure~\ref{fig:graphb} shows a network prober targeting only the MySQL server
port in addition to two of its alternatives.  Figure~\ref{fig:graphc}
corresponds to a network prober targeting remote access services such as SSH (22)
and its alternative (2222), and telnet (23) and its alternative (2323).
Many other probing patterns were identified but they are too large to fit in
this paper.

\begin{figure}[!t]
  \centering
  \includegraphics[scale=.4]{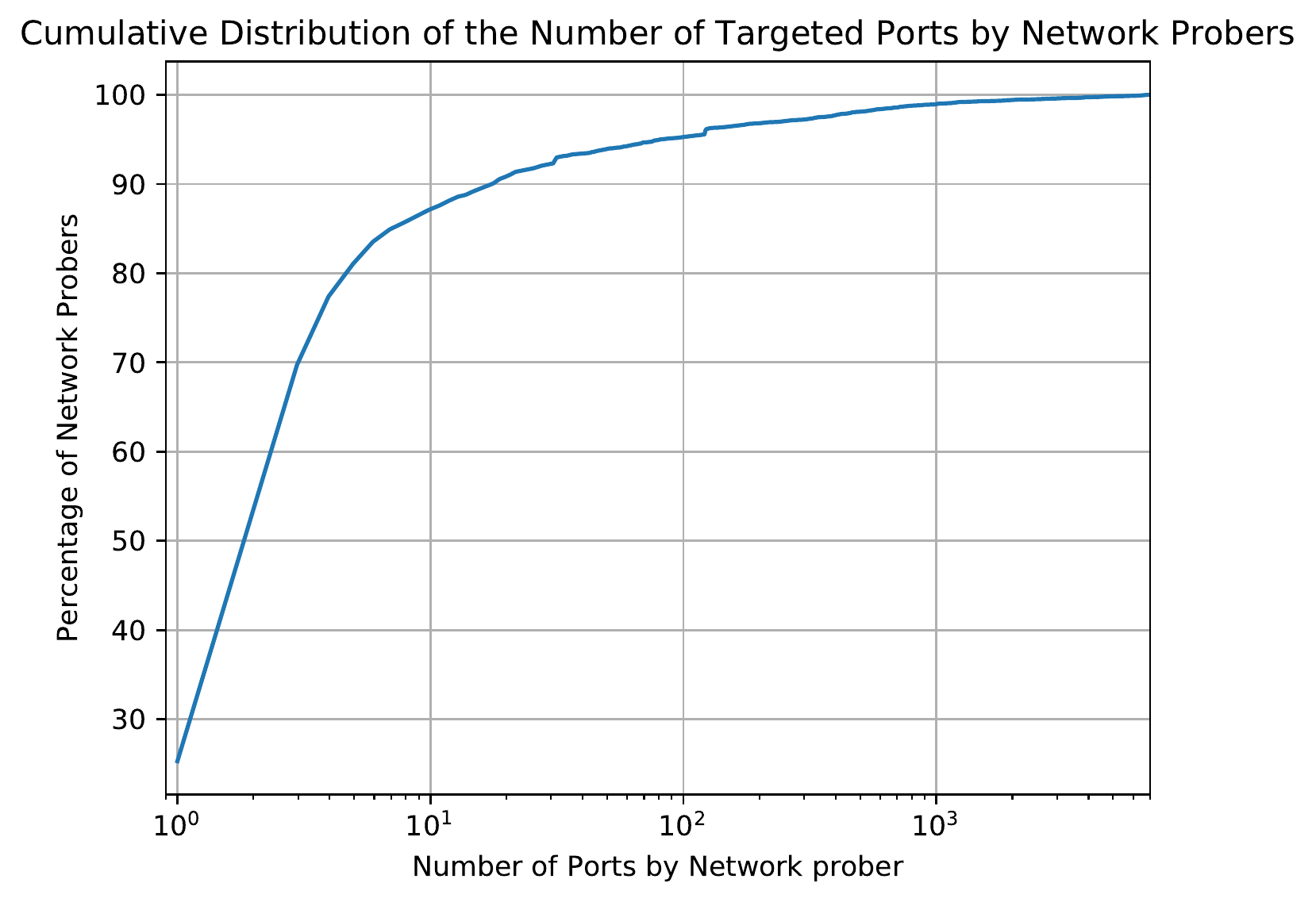}
  \caption{Cumulative distribution of the number of ports targeted by network probers}
  \label{fig:nodes_distribution}
\end{figure}

The extracted transition graphs differ from each other by two main components:
the number of vertices that corresponds to the destination ports and the number
of edges describing the exploration behavior of a network prober.
Figure~\ref{fig:nodes_distribution} represents the cumulative distribution
function of the number of target ports by individual network probers. We
observe that more than 80\% of network probers target less than five ports in
the whole darknet space. This means that most attackers are focusing their
probing activities only on services of interest, which might be related to a
vulnerability disclosure for example.  

\subsection{Relationship Between Ports}

In this section, we aim to identify the relationship between ports in terms of
transition probabilities. We begin by aggregating by network prober the number
of transitions from a port to another. Then, normalize these counts by the
total number of transitions. We repeat this process for all network probers
combined together and for top network probers (as defined in
section~\ref{subsec:top_network_probers_interests}).  

\begin{figure}[!t]
  \centering
  \includegraphics[width=.35\textwidth]{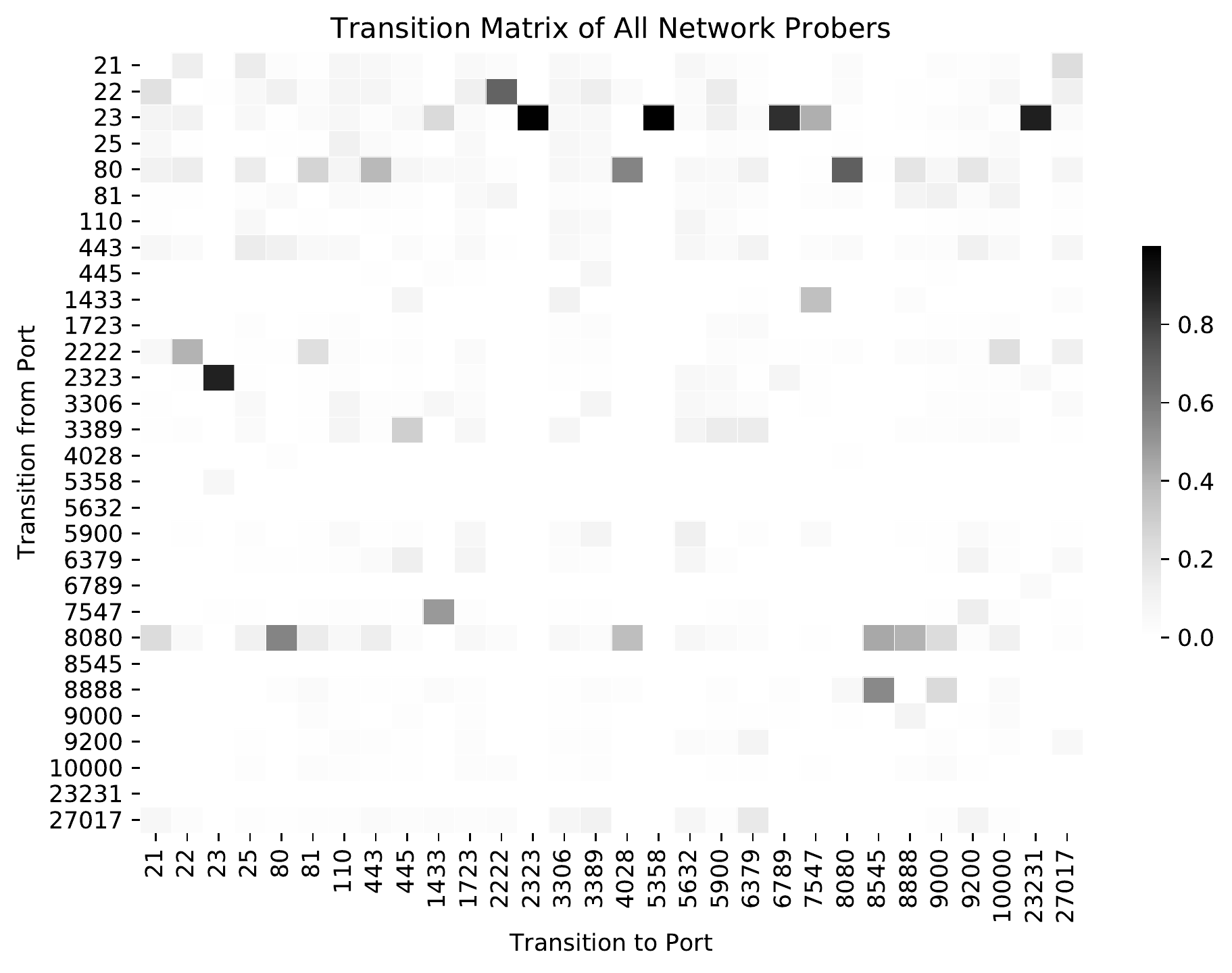}
  \caption{Transition matrix of all network probers}
  \label{fig:matrix_all}
\end{figure}

Figure~\ref{fig:matrix_all} represents the transition matrix of the 30 most
targeted ports in the whole darknet (see
section~\ref{subsec:traffic_by_port}). Unsurprisingly, the figure shows a high
association between ports and their alternatives: 23 and 2323, 80 and 8080, and
22 and 2222. The figure also emphasizes a strong relationship between services
of the same type such as MS-SQL SERVER (1433) and MySQL (3306).

\begin{figure}[!t]
  \centering
  \includegraphics[width=.35\textwidth]{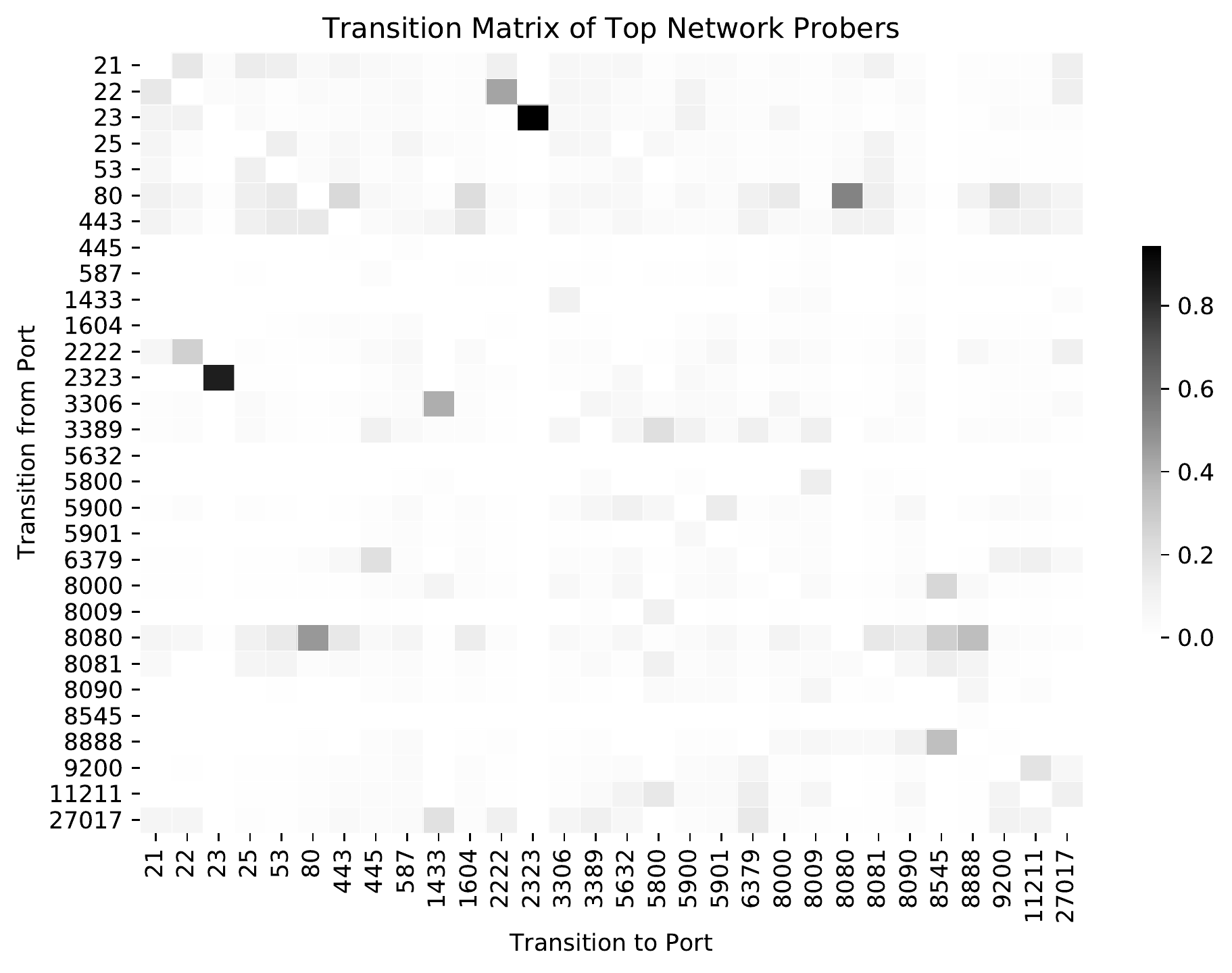}
  \caption{Transition matrix of top network probers}
  \label{fig:matrix_top}
\end{figure}

Similarly, Figure~\ref{fig:matrix_top} shows the transition probabilities of
the 30 most targeted ports by top network probers.  We observe fewer
relationships compared to the previous transition matrix. Nevertheless, the SSH
and telnet services as well as their alternatives still strongly related.

\section{Prediction of Port Probing Rates}
\label{sec:probing_rates}
Predicting probing activities at the service level is a key feature for making
better security operational decisions. Observing a significant disparity
between the predicted probing rate and the actual value may help detecting an
imminent threat. In this section, we forecast the probing rate of a target port
by measuring its previous probing rates as well as the traffic received by the
other ports. The predictions are performed one step ahead of time using the
non-stationary autoregressive model (AR) and the non-stationary vector
autoregressive model (\(\text{VAR}\)), for each port of the 550 most targeted
ports (see Section~\ref{subsec:traffic_by_port}), for different time
resolutions.

\subsection{Data Set}

The probing rate is inferred for different time resolutions: 1 hour, 3 hours, 6
hours, 12 hours and 1 day. For each time resolution, we extract a data set
consisting of the probing rate time series of the most targeted ports
\(\mathcal{D} = \{X^{(i)}\}_{i \in [1, 550]}\) in which each record is a vector
of probing rates occurring in the same time interval. 

\subsection{Non-stationarity of Probing Rate Time Series and Design Parameters}
\label{sec:design_params}

The analysis of the port probing rate time series showed non-stationarity of
the first and the second order statistics over the period of 3 years. However,
we observed that when considering shorter time windows, the non-stationarity
tends to be alleviated at least in terms of average. Therefore, we introduce a
``rolling window'' over the probing rate time series. We train the estimators
using the data falling in the rolling window, then the prediction is performed
one step ahead of time.  The size of the rolling window is a design parameter
that could be interpreted as follows: a short rolling window allows to track
the trend in time while larger rolling windows has the effect of averaging the
trend. Another design parameter to consider is the autoregressive
order which allows the model to infer the linear short term dependencies.

\subsection{Non-stationary AR and VAR Models}

To forecast probing rates, we use the non-stationary autoregressive model of
order \(p\) given by:
\begin{equation}
  \label{eq:ar_model}
  x^{(i)}(t; p) = 
    w^{(i)}_{0}(t) 
    + \sum_{h = 1}^{p} w^{(i)}_{h}(t) x^{(i)}(t - h)
    + \epsilon^{(i)}(t)
\end{equation}
\noindent where \(x^{(i)}(t)\) is the probing rate received by the \(i^{th}\) port at time \(t\).
\(\epsilon^{(i)}(t)\) is the white noise at time \(t\). \(W^{(i)}_t = (w^{(i)}_0(t), w^{(i)}_1(t), \dots, w^{(i)}_p(t))\) is the vector of the model parameters. 
These parameters are estimated using data falling in the rolling window (see Section~\ref{sec:design_params}) and they vary in time.

The non-stationary vector autoregressive model of order \(p\) is given by:
\begin{equation}
  x^{(i)}(t; p) = 
    w^{(i)}_{0}(t)
    + \sum_{j \in \mathcal{I}_k} 
        \sum_{h = 1}^{p}
          w^{(ij)}_{h}(t) x^{(j)}(t - h)
    + \epsilon^{(i)}(t)
\end{equation}
\noindent where \(x^{(i)}(t)\) is the probing rate received by the \(i^{th}\) port (the target port) at time \(t\), 
\(\mathcal{I}_k\) is the set of indexes of the \(k\) retained features (see Section~\ref{sec:feature_selection}), 
\(x^{(j)}(t)\) is the probing rate on the \(j^{th}\) selected port at time \(t\), 
and \(\epsilon^{(i)}(t)\) is the white noise at time \(t\).
\(w^{(i)}_0(t)\) and \(\{w^{(ij)}_{h}(t)\}_{j \in \mathcal{I}_k, h \in [1, p]}\) are the parameters of the model varying in time.

\subsection{Training Algorithm}

After transforming the time series into a supervised learning problem, the
algorithm used to train the non-stationary AR and VAR regressors is the
straightforward normal equation given by:
\begin{equation}
  W^{(i)} = (X^T X)^{-1} X^T X^{(i)}
\end{equation}
\noindent where \(W^{(i)}\) is the vector of the trainable weights, \(X =
(X^{(j)})_{j \in \mathcal{I}_k}\) is the probing rates feature matrix and
\(X^{(i)}\) is the probing rates response vector. In order to reduce the
computation complexity, no regularization is used.

\subsection{Design Parameter Selection}

The design parameters, namely the size of the rolling window \(N\) and the
autoregressive order \(p\), are determined using a grid search strategy. We
varied \(p\) in \([1, 10]\) for the 5 considered time resolutions. Then, we
tried an exhaustive set of rolling window sizes for each autoregressive order
\(p\). The range of \(N\) starts with \(10 \times p\) time units and ends with
75\% of the time series length (leaving thus at least 25\% of data for
validation) with an increment of 10 time units. The optimal design parameter
values \(p^{\star}\) and \(N^{\star}\) are given by the estimator providing the
best coefficient of determination \(R^2\).

\subsection{Feature Selection for the Non-stationary VAR Model}
\label{sec:feature_selection}

To improve the performance of the non-stationary VAR estimators, we select
features according to their individual effect on the response variable using
the Pearson correlation coefficient. This process has the effect of reducing
the noise introduced by uncorrelated features.  First, we split the data
falling in the rolling window (the one giving the best non-stationary AR
performance) into two subsets: a feature selection set \(\mathcal{F}\)
including 75\% of the data and a validation set \(\mathcal{V}\).  Second, we
compute on \(\mathcal{F}\) the univariate correlations in term of probing rates
between the target time series and the time series serving as features to the
non-stationary VAR model, including the autoregressive features.  Third, we
iteratively select the \(k\) most correlated features which we use to train
non-stationary VAR model on \(\mathcal{F}\) and we calculate the coefficient of
determination \(R^2\) on the validation set \(\mathcal{V}\).  The optimal set
of features given by our feature selection strategy is the one providing the
best coefficient of determination.  It is worth mentioning that the selected
features may change over time based on the location of the rolling window in
the time series.  Finally, the selected features are scaled to zero mean and to
unit standard deviation.

\subsection{Results and Discussion}

Table~\ref{table:perf_ar_var} summarizes the performance of the non-stationary
AR and VAR models for 5 different time resolutions for a set of popular
services. It also includes the optimal design parameters for the non-stationary
AR estimators. We used the same design parameters for the non-stationary VAR
estimators.

The performance of the regressors tends to increase for larger time
resolutions, for all the ports except the telnet service.  This is due to
lowered stochasticity of probing rates when considering larger time
resolutions.  Also, we observe that the probing rates of remote access services
are the most predictable. The reason is that such services are highly targeted by
network probers and their probing rate time series are stationary when
considering short time resolutions.

\begin{figure*}[!t]
  \centering
  \includegraphics[scale=0.4]{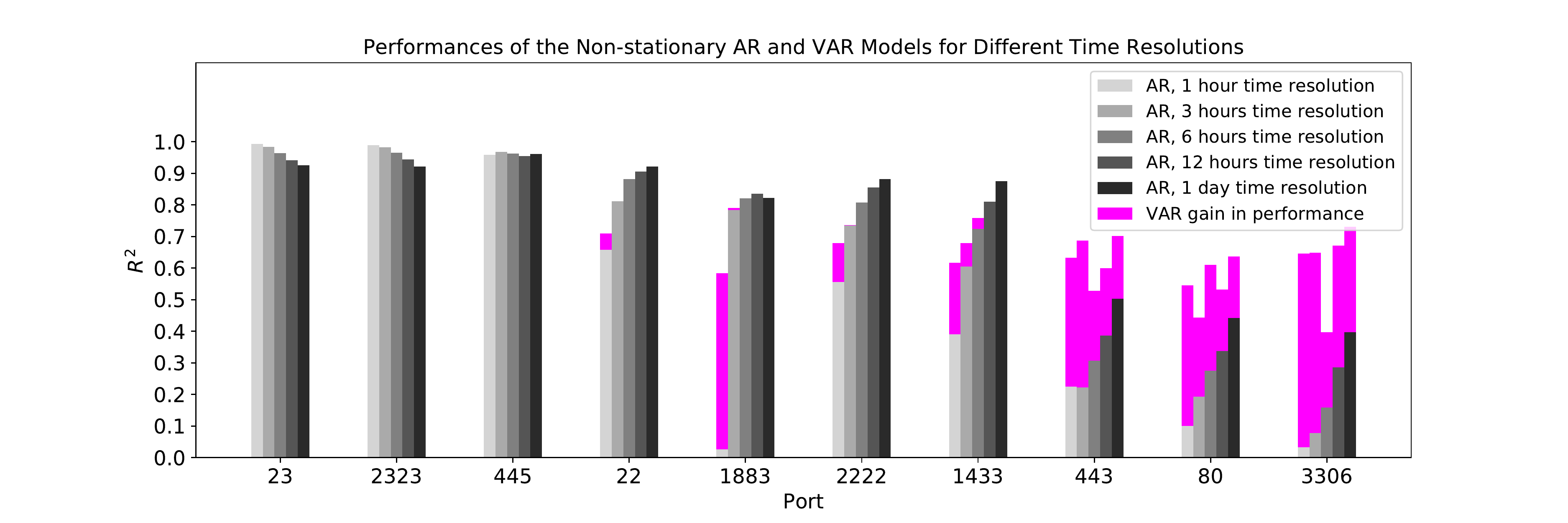}
  \caption{Comparison of performances of the non-stationary 
           \(\text{AR}\) and \(\text{VAR}\) models}
  \label{fig:nsar_vs_nsvar_plot}
\end{figure*}

Also, we observe that the non-stationary \(\text{AR}\) model produces
satisfying results for services exhibiting low short term probing rates
variability such as telnet (ports 23 and 2323).
Figure~\ref{fig:nsar_vs_nsvar_plot} shows that non-stationary \(\text{VAR}\)
model consistently produces better results for services exhibiting high probing
rate variability such as the web services (ports 80 and 443) and the database
management systems (ports 1433 and 3306).

It is noteworthy that the non-stationary autoregressive model fails in
predicting abrupt probing rate changes because of its persistence property.
More powerful and stable models such as FARIMA+GARCH could be used to predict
these extreme values if the probing rate time series exhibit long-range
dependence phenomenon \cite{zhan2013, zhan2015}.  Also, such non-stationary AR
and VAR models, as defined in our paper, require constant update of their
parameters (the trainable weights) and their hyperparameters (the selected
features) due to the non-stationarity of the probing rate time series, which is
computationally expensive. 

\begin{table*}[!t]
  \centering
  \tabcolsep=1.5mm
  \begin{tabular}{|l||rrrr|rrrr|rrrr|rrrr|rrrr|}
\hline
Network & \multicolumn{4}{c|}{1 hour} & \multicolumn{4}{c|}{3 hours} & \multicolumn{4}{c|}{6 hours} & \multicolumn{4}{c|}{12 hours} & \multicolumn{4}{c|}{1 day} \\
Service & $p^{\star}$ & $N^{\star}$ & $R^2_{\text{ar}}$ & $R^2_{\text{var}}$ & $p^{\star}$ & $N^{\star}$ & $R^2_{\text{ar}}$ & $R^2_{\text{var}}$ & $p^{\star}$ & $N^{\star}$ & $R^2_{\text{ar}}$ & $R^2_{\text{var}}$ & $p^{\star}$ & $N^{\star}$ & $R^2_{\text{ar}}$ & $R^2_{\text{var}}$ & $p^{\star}$ & $N^{\star}$ & $R^2_{\text{ar}}$ & $R^2_{\text{var}}$ \\
\hline
\hline
23 (telnet)        &           1 &       17530 &              0.99 &               0.99 &           8 &        5930 &              0.98 &               0.98 &           4 &        2880 &              0.96 &               0.96 &           6 &        1480 &              0.94 &               0.94 &           1 &          60 &              0.93 &               0.93 \\
2323 (telnet alt.) &           4 &       17400 &              0.99 &               0.99 &           2 &        5760 &              0.98 &               0.98 &           4 &        2880 &              0.97 &               0.97 &           1 &         120 &              0.94 &               0.94 &           1 &         720 &              0.92 &               0.92 \\
22 (ssh)           &          10 &       19090 &              0.66 &               0.71 &          10 &        6380 &              0.81 &               0.81 &           9 &        3190 &              0.88 &               0.88 &          10 &        1590 &              0.91 &               0.91 &           9 &         800 &              0.92 &               0.92 \\
2222 (ssh alt.)    &          10 &       19430 &              0.56 &               0.68 &          10 &        6390 &              0.73 &               0.74 &           8 &        3190 &              0.81 &               0.81 &           9 &        1590 &              0.86 &               0.86 &           6 &         800 &              0.88 &               0.88 \\
445 (microsoft-ds) &          10 &       16080 &              0.96 &               0.96 &           1 &         160 &              0.97 &               0.97 &          10 &        2960 &              0.96 &               0.96 &          10 &        1590 &              0.96 &               0.96 &           8 &         740 &              0.96 &               0.96 \\
80 (http)          &          10 &       12890 &              0.10 &               0.55 &           8 &         550 &              0.19 &               0.44 &           8 &        2880 &              0.28 &               0.61 &           7 &        1450 &              0.34 &               0.53 &           7 &         800 &              0.44 &               0.64 \\
443 (https)        &           6 &       18870 &              0.23 &               0.63 &           8 &        6430 &              0.22 &               0.69 &           8 &        3230 &              0.31 &               0.53 &          10 &        1610 &              0.39 &               0.60 &           9 &         800 &              0.50 &               0.70 \\
3306 (mysql)       &           1 &         700 &              0.03 &               0.65 &           1 &         250 &              0.08 &               0.65 &           8 &        2860 &              0.16 &               0.40 &          10 &        1440 &              0.29 &               0.67 &           8 &         720 &              0.40 &               0.73 \\
1433 (mssql)       &           1 &         120 &              0.39 &               0.62 &           1 &          60 &              0.61 &               0.68 &           9 &        2940 &              0.72 &               0.76 &          10 &        1460 &              0.81 &               0.81 &           5 &         730 &              0.88 &               0.88 \\
1883 (mqtt)        &           1 &         360 &              0.03 &               0.58 &           9 &        6000 &              0.78 &               0.79 &           9 &        3000 &              0.82 &               0.82 &          10 &        1510 &              0.84 &               0.84 &           7 &         730 &              0.82 &               0.82 \\
\hline
\end{tabular}

  \caption{Performances of non-stationary AR and VAR estimators for
           different time resolutions for a set of popular services. 
           \(p^{\star}\) and \(w^{\star}\) are the optimal design parameters 
           for the non-stationary AR estimators.}
  \label{table:perf_ar_var}
\end{table*}

\section{Conclusion}
\label{sec:conclusion}
This work presented an exploratory data analysis performed on 2 TB of traffic
collected by a network telescope during the period of 3 years. The
investigation of the network telescope traffic showed that 90\% of probing
activities are targeting only 550 ports of the port space. The latter include
remote access services which are the most sought by network probers, followed
by database management systems, web services and miscellaneous services as
well. This is providing an insight about services requiring particular security
efforts.

The second task was about inferring network probers reconnaissance patterns. We
modeled these exploration behaviors using transition graphs showing the
probabilities of switching from a port to another. This would be exploited in
different applications such as port clustering or classifying network probers
based on their exploration behaviors.

Finally, we assessed to which extent the non-stationary autoregressive and
vector autoregressive models could produce reliable short term probing rate
predictions at the port level. Due to their short memory property, such models
could be used to learn non-stationary probing rate processes with short term
persistence. However, when probing rate processes exhibit long-range 
dependencies, more robust models could be utilized such as GRU and LSTM 
recurrent neural networks.

\section*{Acknowledgment}
This research work is part of the
ThreatPredict\footnote{\url{https://threatpredict.loria.fr}} project partly
funded by the NATO Science for Peace and Security (SPS) programme under
research contract SPS G5319 ``ThreatPredict: From Global Social and Technical
Big Data to Cyber Threat Forecast''.  We acknowledge the support from the
National Center of Scientific and Technical Research (CNRST), Rabat, for the
grant of an excellence scholarship.  The authors would like to thank
Fr\'{e}d\'{e}ric Beck from the High Security Laboratory at INRIA Nancy-Grand
Est, LORIA, for his efforts in managing data and computation servers.

\bibliographystyle{utils/bibtex/IEEEtran}
\bibliography{utils/bibtex/IEEEabrv,src/references,src/references_generated}

\end{document}